\documentclass[11pt, a4paper, titlepage]{article}
\usepackage[left=2cm, right=2cm, top=2cm, bottom=2cm]{geometry}
\usepackage{graphicx}
\usepackage{amsmath}

\usepackage{tocloft}
\usepackage{pdfpages}
\usepackage{array}
\usepackage{multirow}

\usepackage{caption}
\usepackage{booktabs}
\usepackage{rotating}
\usepackage[hidelinks]{hyperref}

\usepackage[style=authoryear,citestyle=authoryear,dashed=false,doi=false,isbn=false,url=false,eprint=false,maxbibnames=8,minbibnames=6,maxcitenames=2,backend=bibtex]{biblatex}
\DeclareDelimFormat{nameyeardelim}{\addcomma\space}
\DeclareFieldFormat{journaltitle}{\textbf{#1}}
\DeclareFieldFormat{journaltitle}{#1}
\DeclareFieldFormat*{title}{\textbf{#1}}
\DeclareFieldFormat{title}{#1}
\DefineBibliographyStrings{english}{%
	andothers = {\em et\addabbrvspace al\adddot}}
\renewbibmacro{in:}{}
\usepackage{caption}
\begin{document}

\section*{Root/Additional Metric (RoAM) framework: a guide for goal\-centred metric construction}

\bigskip

\large{Luke E. B. Goodyear$^{1}$* and Daniel Pincheira-Donoso$^{1}$*}

\bigskip
\bigskip

\noindent\small{\textsuperscript{1}MacroBiodiversity Lab, School of Biological Sciences, Queen’s University Belfast, 19 Chlorine Gardens, Belfast, BT9
5DL, United Kingdom.

\noindent *Corresponding authors: L.E.B. Goodyear, e\-mail: lgoodyear01@qub.ac.uk; D. Pincheira\-Donoso, e-mail: D.Pincheira\-
Donoso@qub.ac.uk, Phone: +44(0)28 9097 2967}

\bigskip
\bigskip
\normalsize

\noindent
\textbf{ABSTRACT} 

\bigskip

The use of metrics underpins the quantification, communication and, ultimately, the functioning 
of a wide range of disciplines as diverse as labour recruitment, institutional management, 
economics and science. For application of metrics, customised scores are widely employed 
to optimise progress monitoring towards a goal, to contribute to decision-making, and to 
quantify situations under evaluation. However, the development of such metrics in complex 
and rigorous settings intrinsically relies on mathematical processes which are not always 
readily accessible. Here, we propose a framework for construction of metrics suitable for 
a wide range of disciplines, following a specified workflow that combines existing decision 
analysis and utility theory concepts to create a customisable performance metric (with 
corresponding uncertainty) that can be used to quantitatively evaluate goal achievement. 
It involves dividing criteria into two groups (root and additional) to utilise a newly 
proposed alternative form of utility function designed to build such customised metrics. 
Once the metrics are produced by this approach, these metrics can be used on a varied set 
of contexts, including their use in subsequent statistical analysis with the metric values 
as a response variable, or informing a decision-making process. The flexibility of the metric 
construction makes it suitable for a wide range of fields and applications, and could provide 
a valuable first step for monitoring and comparison in many different settings.

\bigskip
\bigskip

\noindent\textbf{KEYWORDS}

\emph{metric, methodology, decision analysis, utility theory}

\newpage
\section{Introduction}

The development of methods to calculate metrics representing real-world phenomena 
underpins the assessment, communication and dynamical functioning of political, 
economic and scientific disciplines. Metrics offer common units to objectively assess 
the expression of phenomena that can have a variety of causes (e.g., tons of carbon; 
body mass index - BMI), and enable the opportunity to make standardised comparisons 
across groups or scientific fields. The assessment and study of certain domains can 
fundamentally rely on the availability of some widely established metrics, such as the 
Intelligence Quotient (IQ) in a range of areas aiming to account for variation in cognitive capabilities 
\parencite{Lacalle2023,Makharia2016}, or the Body Mass Index (BMI) in public health 
\parencite{Nuttall2015,Nihiser2007,Cole1995}. Many such metrics are ratio-based, 
such as the BMI or the widely influential journal impact factor in scientific publishing 
\parencite{Clarivate1994}, the units of which relate to something practical 
(kg/m\textsuperscript{2} for BMI, and number of citations per publication for the most recent two year period for 
journal impact factor). However, some metrics are less concrete or remain under constant scrutiny and 
debate, such as the Human Development Index, which is the geometric mean of three 
separate metrics (life expectancy index, education index and income index) and was 
constructed to be indicative of a country’s level of development 
\parencite{Anand2000,Lind2019,Programme2025,Sagar1998}. These types of metrics 
are the focus of this work. Metrics that have total flexibility when it comes to 
the inclusion of contributing factors, but centre around achieving a goal, denoted by a maximal value.

Metrics also play a central role in the functioning and development of societies. 
The settings can be straightforward. For example, in human resources, candidates are 
often graded across several recruitment stages, such as written application, experience and 
interview. These stages are weighted as to their importance in the selection process, and 
the sum of stage grades multiplied by stage weights give final candidate scores. However, in more 
complex settings or when aiming to assess a more complex situation, 
the construction of metrics can be difficult given that it is often necessary to incorporate 
robust mathematical processing, such as decision analysis and utility theory, in the underlying 
calculation, and, can therefore result in some designs that are not as fit-for-purpose as would be 
desired, or even the absence of a much-needed metric.

Here, we propose a methodology, the Root/Additional Metric (RoAM) constructor, which integrates 
different existing methodologies\textemdash decision and utility theory\textemdash into a single workflow to improve 
accessibility of such methods to the wider research community. The RoAM constructor details the use of 
a form of utility function to construct a weighted metric (the Root/Additional Metric (RoAM)) 
that is inherently tailored to the needs of users. With the inclusion of a custom comparison rubric, 
any data can be included that shares the same overarching goal, even if those data contain different 
reporting methodologies. As well as existing methods, such as criteria and criteria weight selection 
from decision theory, this guide includes the following specifications and enhancements which are 
either novel or have not before been applied under the logic proposed here: (i) the splitting of 
criteria into root and additional, and hence the ensuing construction of a RoAM; (ii) incorporation 
of data with different reporting methodologies; (iii) calculating uncertainty for metric values using 
chosen uncertainty variables to construct effective sample sizes; and (iv) the recommendation of the 
construction of RoAMs to be used as measures across a range of technical and scientific disciplines, 
as well as social sciences, economics among others, where constructed metrics are necessary to describe 
broad concepts (e.g., as success in conservation interventions, extinction risk) and to understand the 
relationships between these concepts and pertinent variables. After the construction of the metric, 
it can be used for any downstream requirements, including in statistical analysis and for decision 
making purposes. To outline the mathematical rationale of the RoAM constructor, we first 
describe the full workflow, and then present a simple example, showing a possible practical 
application of the RoAM constructor with a hypothetical scenario, generated to showcase the 
flexibility of the method. Overall, this guide should be able to be applied to any field that 
requires an artificial metric to represent complex data or concepts, offering a versatile and 
generalised methodological approach. 

\section{The RoAM framework}

The framework presented here is divided into three stages: (1) metric construction; 
(2) uncertainty calculation; and (3) further analysis. Stages (1) and (2) are further 
subdivided into (a) variable selection and evaluation; (b) defining variable weighting; 
(c) combining variables and weights to construct metric and uncertainties (see Fig. 1).  
Stage (3) is not described in any detail here but refers to the next steps that may be 
taken once a metric is constructed.

\begin{figure}[b!]
	\centering
	\includegraphics[width = 0.95\hsize]{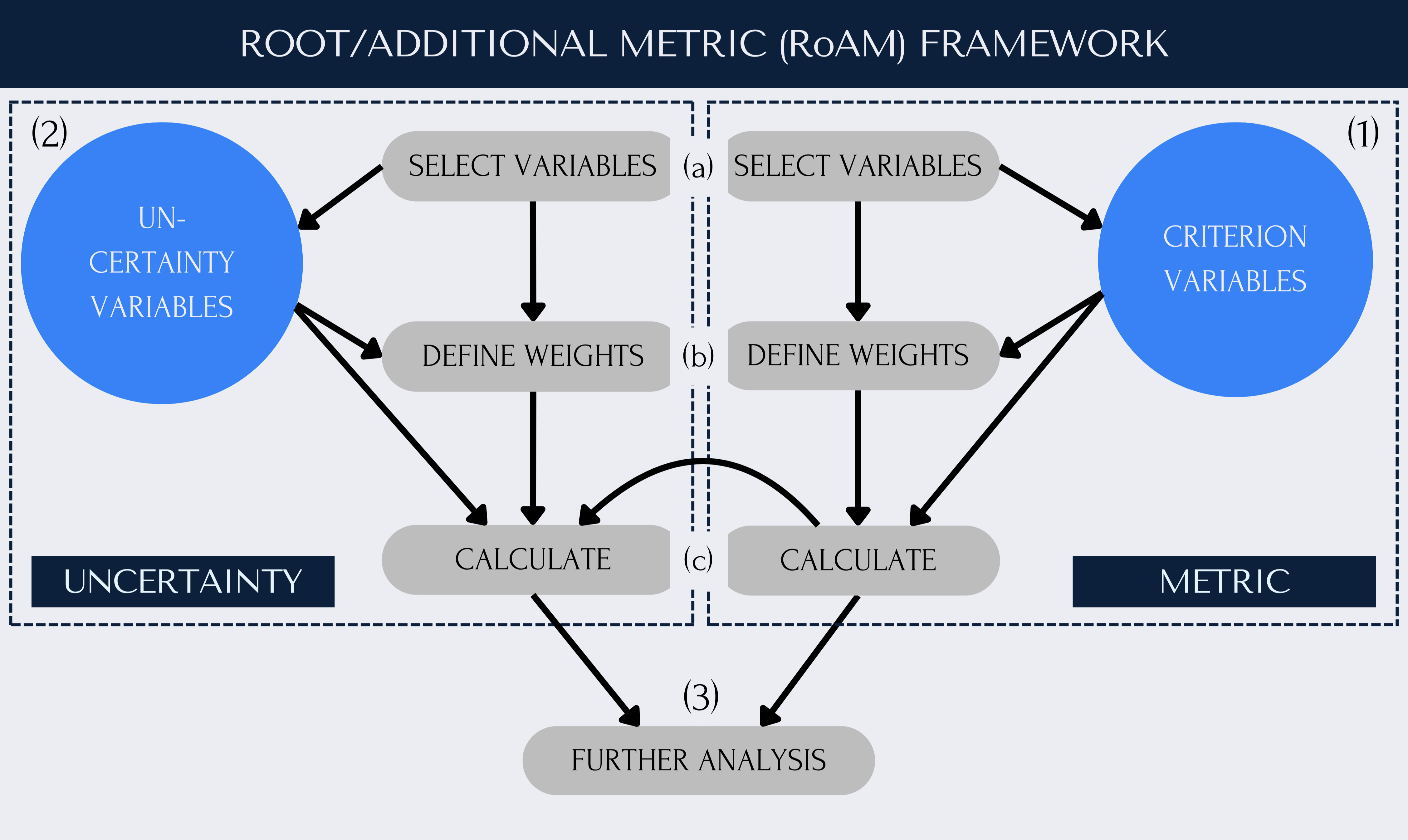}
	\caption{\textbf{Graphical representation of RoAM framework.} 
    The framework is divided into three sections: 
    (1) metric construction; (2) uncertainty calculation; and (3) further analysis. 
    (1) and (2) are further subdivided into 
    (a) variable selection and evaluation; (b) defining variable weightings; 
    and (c) combining variables to construct metric and calculate uncertainties. 
    Only steps (1) and (2) are described in detail in this work, 
    since step (3) can be any method of analysis, such as statistics, 
    or any further use, such as to make decisions.}
\end{figure}

\subsection{Metric construction}

\paragraph{Select criterion variables \\ \\ }

The RoA metric is constructed using criterion variables, which are selected by the constructor based 
on their requirements. Criterion variables are anything that the constructor wants to contribute 
towards evaluating data. Examples could include measures of efficacy, side effects and cost for drug 
treatments, or test scores and teaching hours required for an educational intervention. All criteria 
must contribute in some way towards achieving (or hindering achievement of) the goal, and therefore 
must be ordinal, e.g. have a direction of preference, such that some values are better than others.
If the data for a single criterion is reported heterogeneously, i.e., using different reporting methodologies, 
these will need to be homogenised. There is no one fit solution here and we advocate a customised approach, 
such as the creation of  a rubric using sector knowledge. The only requirement for quantities to be comparable 
is that they reflect the same qualitative information, despite being affiliated with a different reporting 
measure, since datasets should all share the same overarching goal. In other words, all measures contribute 
to the overall goal but are different ways of reporting on it.

If all reporting measures for a given criterion are continuous variables, these can be adjusted to 
be directly comparable. Importantly, however, this is not the same as looking for statistical 
relationship between reporting measures. Statistics could only be used to infer this if this 
information is available within the same study, i.e., a study uses both measures. Instead, 
the constructor should consider their goal and which relationship between reporting measures is 
most appropriate to to their requirements of meeting that goal.

If any of the reporting measures are ordinal or categorical, a continuous scale cannot be used and 
grades will need to be assigned to enable comparison. The simplest way is to bin items according to 
certain compatible thresholds. Any rubric can be chosen or created that will grade across trials 
and can be tailored to the needs of the constructor.
	
Chosen criteria, once homogenised across data-points, can be in the form of continuous variables 
(scaled between 0 and 1), or ordered categorical variables (scaled between 0 [worst] and 1 [best]), 
including binary variables. If all variables are ordinal, then there will be no outliers to skew the 
distribution of the metric value. However, a constructor using continuous variable criteria should 
check for outliers before scaling. 
	
Any scaling methods can be applied, as long as the resulting variables are scaled between 0 and 1. 
For example, the standard min-max formula for equal space scaling, or, to more heavily weight higher 
categories, uneven spacing may be chosen so that the lower categories are more heavily penalised. 
Finally, for any variables where a higher value implies a less desirable option (such as cost), 
the scaled value must be taken away from 1, as this will ensure all variables are scaled so that 
1 implies the maximal metric value. An example of this can be seen in Equation (\ref{eq:minmax}), using the 
min-max formula, where $x_{i}$ denotes the $i$-th data-point and $x^{j}$ denotes the $j$-th criterion:

\begin{equation}
    \label{eq:minmax}
    \text{Scale}(x_i^j) \text{\;} = \text{\;} \displaystyle\dfrac{x_i^j - \text{min}(x^j)}{\text{max}(x^j) - \text{min}(x^j)}
\end{equation}

\paragraph{Define criterion weighting \\ \\ }

The chosen criterion are then split into two groups: 'root' (R) and 'additional' (A). 
The root group represents criteria that are essential to the metric (only dependent on each other) 
and the additional group represents criteria that are non-essential but desirable 
(only dependent on the root criteria) (see Table 1).

In practice, this means that the metric value will equal zero (true failure) if any of the R 
criterion take the value of zero, and that A criteria result in the scaling of the metric 
value and have the same combined weighting as a single root criterion. Therefore, when evaluating 
our criteria weights, only weights for additional criteria need to be evaluated.

Weights for the $n$ additional criteria, $\beta_{1}\dots\beta{n}$, are chosen based on the preferences 
of the constructor, taking into account their particular objectives and the possible practical applications. 
Additionally, a weight, $\beta_{0}$, must be chosen to reflect the maximal metric value an object 
can achieve if all chosen additional weighted criteria are equal to zero. All weights must be subject 
to the following constraints:

\begin{equation}
    \label{eq:betaspos}
    \beta_0, \text{\;} \beta_1, \text{\;} \beta_2, ...\text{\;} \beta_{n-1}, \text{\;} \beta_n \text{\;} > \text{\;} 0
\end{equation}

\begin{equation}
    \label{eq:betasum}
    \beta_0 + \displaystyle\sum\limits_{i=1}^{n} \beta_i \text{\;} = \text{\;} 1
\end{equation}

These can be assigned manually, or by following an existing decision analysis algorithm \parencite{Ezell2021}.

\paragraph{Combine criteria to calculate RoAM values \\ \\ }
To combine criteria into a single performance metric, we are proposing an alternative form of the utility 
function, centred on the division of criteria into root and additional, as described in the previous section. 
This alternative function combines the additive and multiplicative utility models \parencite{Keeney1974} 
in a way most suitable for tailor-made metric construction, while also decreasing the number of necessary 
terms relative to the full multiplicative model. This Root/Additional (RoA) utility function, 
or RoA metric, is shown in Equation \ref{eq:roam} as a vector operation for a single data entry, 
where $x^{j}$ denotes the $j$-th criterion, $\beta_{j}$ denotes the weight of the $j$-th additional 
criterion and $\beta_{0}$ is our constant representing maximal utility, or metric value, 
when no additional criterion are met. For N criteria, we are assuming the first K-1 are additional 
and K to N are root criteria. 

\begin{equation}
    \label{eq:roam}
    \text{RoAM} \text{\;} = \text{\;} \left(\beta_0 + \displaystyle\sum\limits_{j=1}^{K-1} \beta_j x^{j}\right) \cdot \displaystyle\prod_{j=K}^{N} x^{j}
\end{equation}

\begingroup
    \footnotesize
    \renewcommand{\arraystretch}{1.5}
    \begin{table}[]
        \begin{tabular}{p{2cm} p{3cm} p{3cm} p{3cm} p{4cm}}
            \hline
            Criterion type & Definition & Dependency & Weights & Example \\
            \hline
            Root (R) & Essential properties, whereby if a condition for any one of the R criteria is not met, the metric will necessarily be zero. & Only dependent on other root criteria. & Equal to 1. All of equal weight to each other. & \begin{tabular}[t]{@{}p{4cm}@{}}Job experience:\\ For a senior position, it could be decided that a candidate with no relevant experience may automatically receive a value of 0 for the metric value, regardless of other criterion values. \\ \\ Negative side-effects:\\ If maximum threshold is reached (e.g. dosage of a drug is high enough to cause death of a patient), overall metric value should be 0.\end{tabular} \\
            \hline
            Additional (A) & Properties that are non\-essential but desirable, whereby the meeting of conditions will increase the value of the metric & Only dependent on root criteria. & Weights chosen by constructor for each criterion. Combined sum of additional criteria is of equal weight to root criterion. & \begin{tabular}[t]{@{}p{4cm}@{}}Car mileage:\\ For buying a second-hand car, high mileage may be overlooked if the condition is good and the price is low.\\ \\ Conservation long-term impact:\\ If beneficial effects of an conservation intervention are long-term, metric value should increase, however no long-term effects does not mean intervention was a complete failure (equivalent to metric value of zero)\end{tabular} \\
            \hline
        \end{tabular}
        \caption{\textbf{Description of the two types of criteria.} 
        Criteria are designated as root or additional, reflecting their dependencies 
        and impact on the value of the metric, according to the definitions in the table.}
    \end{table}
\endgroup

\subsection{Uncertainty}

Uncertainty can be calculated in either of two situations: (i) for any data-point that is an 
aggregate, i.e., is a summary of a more fine-grained dataset that we do not have access to, 
and therefore contains a sample size; (ii) for any data that contains replicates (e.g., 
data-points that could be aggregated). We also use uncertainty to refer to two different 
but related measures: standard error and uncertainty weights.

Uncertainty weights are scaling factors between 0 (lowest weight) and 1 (highest weight), 
which can be in weighted regression analyses, if a user of a RoA constructed metric aims to 
perform statistical analyses after calculating metric values. These serve the purpose of 
ensuring that data points with more certainty around their metric value have more influence 
on the regression coefficients. For example, sample size is included in the calculation, 
reducing the weight of data-points with small samples.
	
Standard errors use the same uncertainty variables (e.g. sample size), but are calculated 
using an estimation of a beta distribution. Here we treat the metric value as the mean of a 
beta distribution, and calculate the standard deviation to simulate uncertainty. This 
allows us to artificially construct a beta distribution based on each data-point as if 
it were an aggregate of multiple data-points. If we have access to replicates or the original, 
non-aggregate dataset, we may model the distribution directly, calculating metric values for 
each of these data-points and so calculate the distribution of the metric value directly. 
From the standard errors, we can then calculate confidence intervals.

\paragraph{Select uncertainty variables \\ \\ }

Uncertainty variables are selected in the same way as criteria, based on constructor 
requirements. Uncertainty variables can be any variable that contributes to uncertainty 
around the metric value, such as a measure of usability representing clearness/robustness 
of methodology and data. However, sample size is always included, which is the only fixed 
requirement for the whole workflow.

\paragraph{Define uncertainty variable weightings \\ \\ }

As with criteria, all uncertainty variables must be scaled between 0 and 1, where 0 
is high uncertainty/low precision, and 1 is low uncertainty/high precision. However, 
since all uncertainty variables are multiplicative, the entire uncertainty weight is 
directly down-weighted by each uncertainty variable. 

Sample size can be scaled using any function that transforms positive integers 
to a value between 0 and 1,  depending on the dataset. If sample sizes are all the 
same, or equal weighting is required, this value can just be set to a constant of 1. 
A linear scale should be adequate for sample sizes that are no more than a multiple 
of ten between the smallest and the largest. For extremely variable sample sizes, such 
as a range that covers many values on a $log_{10}$ scale, we recommend setting a capping 
threshold, whereby any sample size above this is given a weighting of 1, and using the 
logistic function for scaling (Appendix 3.1). 

For all other uncertainty variables, any appropriate function may be used, but 
a simple method would be to pick the weight for the lowest grade (which would 
become the minimum weight of any data-point with reference to that particular 
uncertainty variable, and hence referred to as the minimum threshold for the variable) 
and calculate a linear function between this lowest weight and 1, distributing the grades evenly.

\paragraph{Combine uncertainty variables to calculate uncertainties}
\subparagraph{Uncertainty weights \\ \\ }

Calculating uncertainty weights is much simpler than calculating metric values as all variables are multiplicative. Therefore, all uncertainty variables are multiplied together with the scaled sample size:

\begin{equation}
    \label{eq:uncertaintyweights}
   \text{Uncertainty} \text{\;} = \text{\;} \text{S} \cdot \displaystyle\prod_{j=1}^{N} U^{j}
\end{equation}

where $S_{i}$ is the sample size of aggregate data-point $x_{i}$ and $U_{i}^{j}$ is the 
value of the uncertainty variable $j$ for data-point $x_{i}$.

\subparagraph{Standard errors and confidence intervals \\ \\ }

Calculating standard errors uses the standard deviation formula for a beta distribution. 
We can construct an estimation of a beta distribution for each aggregate data-point using the
$\mu/\nu$ parametrisation, where $\mu$ is location and $\nu$ is shape. 
For the RoAM framework, $\mu$ is just our metric value, which becomes the mean of 
our constructed beta distribution, and $\nu$, which is, strictly speaking, 
the sum of the probabilities of each outcome (0/1) and is related to the sample size, 
can be calculated in a similar way to our uncertainty weights, except we use the raw 
sample size instead of the capped scaled sample size. This means that $\nu$ acts as an 
effective sample size, weighted by the uncertainty variables. we refer to this generated 
beta distribution as an artificially constructed from aggregate (ACA) beta distribution to, 
firstly, differentiate it from the beta distribution created by the metric values of the 
dataset that may be required for a  statistical analysis stage, and, secondly, to highlight 
the use of effective sample size, rather than a more mathematically strict calculation of 
$\nu$, and, thirdly, to emphasise the fact it has been artificially constructed from summary data. 
The standard deviation for this ACA beta distribution is then used as an estimate of the standard error 
around the metric value for the aggregated data. This means that each aggregate data-point has its own 
ACA distribution, from which we can make inferences about the original fine-grained dataset without 
having any access to it. 
In other words, each data-point is thought of as its own beta distribution (since it is an aggregate 
of a whole dataset), with its metric value as the mean and its standard deviation calculated using 
effective sample size. Since the distribution is necessarily bounded between 0 and 1, precision 
increases towards 0 and towards 1, which means variance is related to mean, as expected for a 
beta distribution. This makes sense because a value very close to zero is as good as zero and 
a value very close to one is as good as one, in terms of an indication of performance. 
However, error, even at exactly 0 (or 1), should still not be 0, and so the tail from 0 
(or 1) should still be related to effective sample size. This is rectified by adding a small 
value to every 0 and subtracting the small value from every 1 when calculating the error. As 
sample size increases, precision also increases, shrinking the error and confidence interval 
towards this tiny value, which for the purposes of seeing how close we are to reaching the 
desired goal, is our effective zero (or 1) since no mean is below (or above) this value.

The complete general calculation can be seen in Supplementary 1, and provides details of the 
conversion of the $\mu/\nu$ parametrisation into $\alpha/\beta$ parametrisation, the required modification of 0 and 
1 values, and the standard formula for standard deviation of a beta distribution, based on the 
binomial formula. 

Using these values (the mean and standard deviation of each ACA distribution), 
confidence intervals can be created around each of our metric values.

\subsection{Further analysis}

As it has been shown above, the RoAM workflow incorporates utility theory and decision theory to 
weight criteria, and then use these weighted criteria to construct a metric that provides a 
performance indicator value between 0 and 1. These metric values could thus now be used 
(either the mean values or lower and upper confidence limits) as a response variable in further analyses.

\section{Hypothetical worked example: RoAM in action}

To demonstrate the operational rationale and the functionality of RoAM, this section presents a worked 
example based on a hypothetical analysis relevant to conservation biology. Therefore, the variables 
will be described using terminology from the perspective of conservation management to illustrate how 
the RoAM constructor works in practice.

To unfold this hypothetical worked example, we want to construct a metric to describe the success 
of hypothetical conservation management interventions implemented with the goal of mitigating the 
impacts of human industrial activity (e.g., represented in prevalent pressures, such as habitat 
degradation, climate change, tourism, overexploitation of natural resources) on the erosion of 
biodiversity. These could be any form of intervention, for example the creation and management 
of protected areas (e.g., national parks for legal protection of wildlife) or connecting corridors 
(e.g., connections between habitat parts that have been partitioned as a result of industrial activity, 
such as creation of roads). The aim is to assess the different interventions for repeatability in the future.

\subsection{Metric construction}

\paragraph{Select criterion variables \\ \\ }

Criterion variables characterise our aims in achieving maximal metric value, and are chosen 
based on our goals, informed by the main reporting measures in our data. In our case, this 
is improving biodiversity. From the data we have available for each intervention, 
we need to choose which variables should contribute to our metric.

\begin{table}[b!]
    \footnotesize
    \renewcommand{\arraystretch}{1.5}
    \begin{tabular}[l]{>{\raggedright\arraybackslash}p{3cm} | >{\raggedright\arraybackslash}p{2.3cm} >{\raggedright\arraybackslash}p{2.3cm} >{\raggedright\arraybackslash}p{2.3cm} >{\raggedright\arraybackslash}p{2.3cm} >{\raggedright\arraybackslash}p{2.3cm}}
                                        & \multicolumn{5}{c}{Grade}                                                                           \\
                                        & 0             & 1                  & 2                    & 3                 & 4                   \\
        Reporting measure               & Not effective & Slightly effective & Moderately effective & Highly effective  & Extremely effective \\
        \hline
        Single species abundance        & Decreases     & Does not change    & Increases by 20\%    & Increases by 35\% & Increases by 50\%   \\
        Species richness                & Decreases     & Does not change    & Increases by 10\%    & Increases by 15\% & Increases by 30\%  
    \end{tabular}
    \caption{\textbf{Sample rubric equating different reporting measures.} 
    Three different reporting measures are converted to a standardised grading system to enable direct comparison.}
\end{table}

\subparagraph{\emph{Effectiveness}}

The most important criterion is the measure of effectiveness. However, for our hypothetical example, 
this is not measured using a consistent reporting methodology, with two different measures being used: 
change in species abundance for a single species and change in species richness. A sample rubric 
(Table 2) grades each intervention trial according to five different categories 
(not effective, slightly effective, moderately effective, highly effective, and extremely effective), 
corresponding to different thresholds for each of the two reporting measures, enabling straight-forward 
comparison across all categories. We then need to scale this so that our effectiveness grades are between 
0 and 1, which we will do using the standard min-max formula.

\subparagraph{\emph{Cost}}

The total monetary cost of implementation should also factor into the metric, since we want to 
consider a trade-off between favourable outcome and cost. It is a continuous variable and can be 
included by using the min-max formula in Equation \ref{eq:minmax} since higher cost should be penalised more 
than lower cost. For a situation where there is a strict budget cap, a tailored version of this 
formula can be used, with the available budget as the maximum, instead of extracting it from the data, 
returning a value of zero if the cap is breached (negative metric values are always rounded to 0).

\subparagraph{\emph{Politics}}

The versatility of the RoAM approach makes it suitable for the incorporation of additional criteria 
that may not be immediately obvious as factors involved in the process of biodiversity loss. That is, 
further criteria that can impose constraints or advantages and contribute to the difficulty level of 
application, and thus, to the optimisation of the process. For example, an underlying factor involved 
in the implementation of conservation management interventions can be the national political 
conditions\textemdash national prosperity (e.g., economic growth, socio-economic sustainability, 
peace, institutional efficiency) are directly associated with the degree of democratic tradition of a 
country \parencite{acemoglu2012why,acemoglu2019narrow,Stiglitz2022}. The sustained failure of 
institutional functioning is known, in turn, to translate into challenges to the implementation of 
conservation interventions \parencite{Daskin2018}. Therefore, the interventions can involve 
significant degrees of policy and political conditions. When constructing the metric, which 
represents the success of an intervention, we may want to consider ease of implementation in 
terms of this political background, since it gives an indication of repeatability in the future. 
In our scenario, we designate a smooth political process as 1, and give 0 for a complex political 
process, or for missing information, thereby treating absence of consideration of political process 
as a worst case scenario when data is missing. 

\paragraph{Define criterion weighting}

\subparagraph{\emph{Effectiveness}}

As effectiveness is the most important quantity, being the measure of whether the intervention 
has the potential to work at all, it must be a fundamental criterion. This will return the desired 
behaviour, i.e., if effectiveness is zero, we want our metric to equate be zero since this indicates a 
total failure of the intervention.

\subparagraph{\emph{Cost}}

Cost could be included as either a fundamental of additional criterion, depending on strictness of budget cap. 
However, including it as a fundamental criterion could mean you lose valuable data if, for example, you later 
accessed more funding, or an improvement to that trial was found to reduce costs. Therefore, it is recommended 
to use cost as an additional criterion, setting its weight proportionately to the strictness of your budget. 

\subparagraph{\emph{Politics}}

For politics, having a political hoop to jump through does not mean the option failed to work, and we 
are not adverse to working hard to removing red tape if the reward justifies the means, i.e. if effectiveness 
is high enough and cost is low enough to offset this extra work. Therefore, a zero score would not equate to 
zero metric value, and thus, politics is included as an additional criterion.
	
For simplicity, and since we have presented a straight-forward example with only two additional 
criteria plus baseline, weights can be chosen to be allocated manually. Here, we will choose to set 
the baseline (the maximum value for trials that don't meet the additional criteria) as 1/2, 
then, valuing cost three times as much as politics, and ensuring all betas sum to 1, we will get the 
following set of betas:

\begin{align}
\beta_0 &= 1/2 = 0.5 && \label{eq:b0} \\
\beta_{cost} &= 3/8 = 0.375 && \label{eq:bcost} \\
\beta_{politics} &= 1/8 = 0.125 && \label{eq:bpol}
\end{align}

\subsection{Combine criteria to calculate RoAM values}

The metric can be constructed as follows:

\begin{equation}
    \label{eq:exroam}
    \text{RoAM}(x_i) \text{\;} = \text{\;} (0.5 \text{\;} + \text{\;} 0.375 x_i^\text{cost} \text{\;} + \text{\;} 0.125 x_i^\text{politics}) 
    \text{\;} \cdot \text{\;}  x_i^\text{effectiveness}
\end{equation}

\subsection{Uncertainty}

\paragraph{Select uncertainty variables}

\subparagraph{\emph{Sample size}}

In this hypothetical example, say sample sizes have a range of 30 to 200, with no obvious outliers, 
so the cap can be set to the maximum value of the dataset (i.e., 200). Since available sample sizes 
are within a multiple of ten, a simple linear function can be used to scale to between 0 and 1.

\subparagraph{\emph{Usability}}

To calculate uncertainty, we also want to incorporate a variable called usability. 
The hypothetical publications that are being used will not have been written or undertaken 
with this sort of analysis in mind, and thus, there could be pertinent data missing, or methodology 
that does not meet the requirements. Usability is a way of measuring this type of uncertainty, and 
we assign a value from 0 to 3 that corresponds to the clarity and presence of methodology and data, 
where 0 indicates that nothing is lacking and is the maximum score, according to some specified rubric 
we create. We choose to remove any trial with a usability score of 3 simply because the uncertainty is 
too great to estimate a useful metric value.

\paragraph{Definition of uncertainty variable weightings \\ \\ }

Sample size is already scaled appropriately. For usability, the minimum threshold is 
chosen such that this is the weighting applied to a trial with score 2. Again, we will 
do this manually, choosing to give a minimum weight of 0.6, so any trial with a usability of 
2 can only have 60\% certainty as a maximum. The rest of the possible values are distributed evenly, 
giving a final weighting schema of 1 for grade 0, 0.8 for grade 1 and 0.6 for grade 2.

\paragraph{Combining uncertainty variables to calculate uncertainties \\ \\ }

Having achieved scaled sample size and usability weighting, which we can combine to generate our 
uncertainty weights as follows:

\begin{equation}
    \label{eq:exuncert}
    \text{Uncertainty}(x_i) \text{\;} = \text{\;} \text{ScaledSampleSize}_{i} \text{\;} \cdot \text{\;} \text{Usability}_{i}
\end{equation}

We can also combine raw sample sizes with usability weighting, and calculate standard errors as follows:

\begin{align}
\mu_{i} \text{\;} & = \text{\;} \text{Utility}(x_i) && \label{eq:mu} \\
\nu_{i} \text{\;} & = \text{\;} \text{SampleSize}_{i} \text{\;} \cdot \text{\;} \text{Usability}_{i} && \label{eq:nu} \\
a_{i} \text{\;} & = \text{\;} \mu_{i} \cdot \nu_{i} && \label{eq:a} \\
b_{i} \text{\;} & = \text{\;} (1 \text{\;} - \text{\;} \mu_{i}) \cdot \nu_{i} && \label{eq:b} \\
\text{Error}(x_i) \text{\;} & = \text{\;} \sqrt{\dfrac{(a_i \text{\;} \cdot \text{\;} b_i)}{(a_i \text{\;} + \text{\;} b_i)^2 \text{\;} \cdot \text{\;} (a_i \text{\;} + \text{\;} b_i + \text{\;} 1)}} && \label{eq:error} \\
\end{align}

Confidence intervals are calculated the standard way, except with rounding if any 
interval exceeds the inclusive limits of 0 and 1.

\subsection{Next steps}

Here, the user is free to make use of whatever statistical analysis 
techniques they desire, to try and untangle the relationships and biases. Although 
we are not working through this section in our example, since there are many excellent 
guides on how to perform statistical analyses, there is one important consideration to 
re-emphasise. There is a key difference between artificially constructed distributions, 
such as those constructed for each trial, using the metric value as a mean, and raw constructed 
distributions, such as those that find the statistical mean using the summarised dataset. 
The ACA distributions are used only to estimate uncertainty around the calculated metric value 
for each trial. Any other mean or uncertainty is calculated in the usual manner using each trial 
as a data point, incorporating the mean usability score for each groupings, the sum of sample sizes 
across trials, and the mean of the metric value for each grouping. 

\section{Conclusion}

The use of metrics is an important part of monitoring and comparison, and a pervasive element of 
everyday life. To be able to construct metrics that give the information most needed in the most 
succinct format is a valuable tool to achieve any goal.

As such, the RoAM metric framework is a multi-disciplinary method that we hope will be of 
practical use across disciplines. By choosing the most valuable criteria on a subjective basis, 
and dividing these into two groups in order to extract the most fundamental elements, and 
weighting according to requirements, the workflow enables a step-by-step transparent 
construction of any goal-centred metric.

Constructing a metric is a first step in the monitoring of achievement towards a goal, and 
it can be used in subsequent downstream activities to better understand gains and failures, 
explore what is working well and what isn't, or even to generate predictions of which conditions 
may obtain the best performance. These further analyses can make use of any of the multitude 
of highly sophisticated techniques that have been developed over the years, and it is hoped 
that the use of a metric will help to streamline such processes. We also hope that others may 
see the promise in this attempt to generalise the metric construction process, and be able to 
further enhance the technique, making it even more useful to those whom it may benefit.

\newpage

\subsection*{Code availability}

Code for upcoming R package available at \url{https://github.com/lebgoodyear/RoAM}. 
Please see READMEs for requirements and details of use.

\subsection*{Ethics declarations}

The authors declare no competing interests. 

\newpage

\printbibliography

\newpage

\section*{Supplementary 1}

\subsection*{Logistic scaling of capped sample sizes}

For extremely variable sample sizes, such as a range that covers many values on a $log_{10}$ scale, we recommend 
setting a capping threshold, whereby any sample size above this is given a weighting of 1. In this case, 
the logistic function would be a suitable scaling function, as it ensures the biggest difference in effect 
in the mid-range of the $log_{10}$-transformed sample sizes:

\begin{math}
    y \text{\;} = \text{\;} \dfrac{1}{1 + e^{-k (x - x_0)}}
\end{math}

\noindent where $y$ is the new scaled sample size, 
$x_{0}$ is the location of the midpoint of the curve, 
$k$ is a measure of the steepness of the curve 
and, $x$ is the $log10$ transform of the capped sample sizes:

\begin{math}
    x \text{\;} = \text{\;} log_{10}(x_{cap})
\end{math}

\noindent An appropriate choice for $x_{0}$ and $k$ could be:

\begin{math}
    x_0 \text{\;} = \text{\;} \dfrac{x_{max}}{2}
\end{math}

\begin{math}
    k \text{\;} = \text{\;} 10^{-(log_{10}(x_{max})-1)}
\end{math}

\noindent where $x_{max}$ is the $log10$ transform of the maximum capped sample sizes (i.e. the value of the cap):

\begin{math}
    x_{max} \text{\;} = \text{\;} log_{10}(max(x_{cap}))
\end{math}

\noindent This is a specific generalised form of the logistic function, enabling the curve to scale 
with the dataset. It sets the midpoint of the curve to half-way and gives a transition over this central point 
that spans roughly half the dataset.

\end{document}